# Requirements for Ion Sources

*R. Scrivens*
CERN, Geneva, Switzerland

**Abstract**
Ion sources produce beams for a large variety of different physical experiments, industrial processes and medical applications. In order to characterize the beam delivered by them, a list of requirements is necessary. In this chapter the list of principal requirements is specified and definitions for them are given.

## 1    Uses for ion sources

Ion sources are one of the critical components for the production of high-energy beams of particles, which have many applications, not just in the field of scientific studies, but also for medical applications, material modification and surface treatments, and semiconductor doping. Even possibly, in the future, they may find applications in energy production through inertial confinement fusion, magnetic confinement fusion or energy amplification.

As the ion source is a driver for a further process or study, there are certain specifications on its performance, which must be quantified. In this chapter, the definitions of this characterization will be summarized.

## 2    The need for ion sources

One item that all particle accelerators have in common is that they have to start with a source of particles. For many applications and experiments, it is necessary to use heavier particles, such as protons, the nucleus of the hydrogen atom.

In order to impart a large amount of energy to an atom, it is necessary to apply a force to it. We are only able to create electric and magnetic fields in a well-defined manner, and in general the gravitational field is too weak to be able to attain a reasonable energy gain. (For example, a hydrogen atom allowed to free fall to the Earth's surface would reach a total energy of 0.65 eV as it reached sea level.) In order to use the electromagnetic forces, we must give atoms an overall charge. The simplest way to do this is by removing one or more electrons, or attaching additional electrons, and therefore creating ions. The device for doing this is an ion source.

Once transformed into a charged state, the ions can be manipulated with the Lorentz force – see Eq. (1) – which allows them to gain energy and therefore to be accelerated with an electric field, as well as to be deflected with a magnetic field:

$$\vec{F} = qe(\vec{E} + \vec{v} \times \vec{B}). \tag{1}$$

In a more general sense, we consider an *ion source* to be a device that ionizes atoms in order to produce ions, and provides a first stage of acceleration (ion extraction) in order to produce a beam of ions. We define an *ion beam* as an ensemble of ions that move with an average velocity and direction.

## 3    Characterization

The characterization of the ions, and the qualities of the beam into which they are formed, require several definitions. Seen from the viewpoint of the requirements of the user of the ion beam, they are necessary to allow the choice of the type of source to be made.

### 3.1    Particle type

The type of particle required for the application is the most fundamental parameter to be chosen, and is symbolically represented in the following way:

$$\left(^{A}\mathrm{E}^{q+}\right)_{n} \tag{2}$$

The first principal choice is the type of atom from the periodic table (here given the general symbol E for 'element'). Further to this, the application may require a specific isotope to be provided ($A$), which could be one of a stable type, or, in the case of radioactive ion beams, an unstable isotope. The charge state of the ion ($q$) can be either a positive state, up to the fully stripped nucleus, or a negative ion, with the addition of an electron to the atom. Next, the requirements may not be for ionized atoms, but for ionized molecules ($n$), where the atom may be paired with the same atom type, or a molecule could be required of specific, dissimilar atoms (which may have their own isotope specifications).

Finally, the application may require different ion types to be available, which may only be possible with the use of several ion sources, and the time to switch between ion types may be a factor to consider.

### 3.2    Energy

It may be possible to produce ions of the energy required by the application directly from the source itself (and its extraction system), or the ions may need further acceleration in a particle accelerator (using DC or radiofrequency (RF) electric fields). In the case that a further accelerator is needed, the energy from the source must be chosen in collaboration with the subsequent acceleration stage.

The typical definition of the energy of the beam from an ion source is made purely in terms of the kinetic energy, i.e. the rest mass energy is ignored. The energy is then well defined by the voltage used to accelerate the ions, i.e.

$$E_{\text{kinetic, total}} = qeU, \tag{3}$$

where $q$ is the charge state of the ion (1, 2, 3, …), $e$ is the electron charge and $U$ is the applied voltage (or equivalent) used for acceleration. The use of a negative ion requires that the voltage is also reversed. For the definition of the ion energy, the electron charge is usually contained in the units, so a $^{20}\text{Ne}^{5+}$ ion accelerated across 100 kV would be defined as having an energy of 500 keV, whereas in SI units the energy would be $8 \times 10^{-14}$ J.

Many types of RF particle accelerator require a fixed ion velocity at their input. In the case that different ion species are to be delivered by an ion source, and accelerated, it is very common to use the kinetic energy per nucleon as a definition:

$$E_{\text{kinetic, per nucleon}} = \frac{E_{\text{kinetic, total}}}{A} = \frac{qeU}{A} = \frac{1}{2}m_{\text{u}}v^{2} = (\gamma - 1)m_{\text{u}}c^{2}, \tag{4}$$

where $A$ is the mass number of the ion, $m_{\text{u}}$ is the rest nuclear mass unit ($1.67 \times 10^{-27}$ kg), $c$ is the speed of light and $\gamma$ is the relativistic gamma factor. The final two terms demonstrate that the kinetic energy per nucleon is only dependent on the velocity of the particle (either $v$ or $\gamma$).

The advantages of using a higher particle energy are:

- space charge is reduced – higher beam currents are possible;
- reduces energy spread ratio ($\Delta E/E$);
- reduces the geometrical emittance $\Rightarrow$ smaller apertures;
- higher velocity makes it easier to inject into an RF accelerator (e.g. drift tube linac).

And the disadvantages are:

- higher risk of sparking, implying a higher degree of complexity to avoid it;
- longer radio frequency quadrupole (as the input cells become longer);
- higher fields for beam devices;
- higher energy = higher beam power $\Rightarrow$ consequences for beam intercepting devices.

### 3.2.1 *Intensity and repetition rate*

The beam intensity from a source is usually quoted in terms of an equivalent electrical current, i.e.

$$I_{\text{electric current}} = \frac{qeN_{\text{ions}}}{t}. \tag{5}$$

In the case of multiply charged ion beams, the particle current may be given, which is the equivalent current for a beam where all ions have a charged state of +1:

$$I_{\text{particle current}} = \frac{eN_{\text{ions}}}{t} = \frac{I_{\text{electric current}}}{q}. \tag{6}$$

It is normally accepted to quote the beam current during the beam pulse, although this is not always the case.

The repetition cycle must be defined (unless the beam is required in continuous mode). The pulse length and the repetition rate are sufficient, but care must be taken to understand the effects of beam outside this time window.

### 3.3 Space charge

Ions within our beam of particles will all have the same polarity of charge, which means they will all repel each other, an effect that is called *space charge*.

For beams of *electrons*, there is a relatively simple way to characterize how much the space-charge force will affect the beam. This is found by calculating the beam perveance ($P$), which is defined in terms of the beam electric current ($I$) and the beam energy ($U$) as

$$P = \frac{I}{U^{3/2}}. \tag{7}$$

If two electron beams of the same initial beam size and perveance are left to drift through an otherwise field-free region, they will expand at the same rate (relative to the longitudinal direction of travel).

However, this simple perveance equation for electrons is only valid because all electrons are alike. For ions, we have to take into account also the masses and charges of the ions in the beam. In this case, we develop the poissance ($\Pi$), which is defined as

$$\Pi = \frac{9}{4\varepsilon_0} \frac{I}{U^{3/2}} \sqrt{\frac{m}{2qe}} \frac{a}{b}, \tag{8}$$

with the usual definitions for all the symbols, and with *a* and *b* being the height and width of the beam. (This leads to a beam having a different poissance in each plane, with the smaller size having the large poissance value.) The poissance equation allows the comparison of different particle beams, of different energies and ion types. It shows that low-energy, low-charge-state, high-mass beams can have very strong space-charge characteristics; for example, a 40 mA, 50 keV proton beam has the same $\Pi$ as a 0.2 mA, 5 keV Ar$^+$ beam.

An important effect that helps to reduce the space-charge forces of the beam at low energies is the trapping of ions or electrons produced by beam collisions with the residual gas. This space-charge compensation works when the beam is in a DC or quasi-DC mode (i.e. pulsed for some tens of microseconds or more).

### 3.4 Emittance

The emittance of a single particle or a beam can be considered from the viewpoint of the amplitude term of the solution to Hill's equation for the periodic structure of a circular particle accelerator [1]. However, when considering the beam from an ion source, there is no periodic structure with which to compare the ion position, and the emittance definition needs to be considered in a different (but consistent) way.

If we consider a particle in a beam travelling along the beam axis (which we will call the *z*-axis in this case), we can plot the particle's position (*x*) and angle (*x′*) relative to *z* on a graph, which is referred to as a phase-space plot. If this is done for all the particles in a beam, we can draw an enclosing shape around all the particles (see Fig. 1). Then the total, geometric emittance is defined as the area of the phase space covered with particles, divided by π, i.e.

$$\varepsilon_{\text{total, geometric}} = \frac{\text{area}}{\pi}. \tag{9}$$

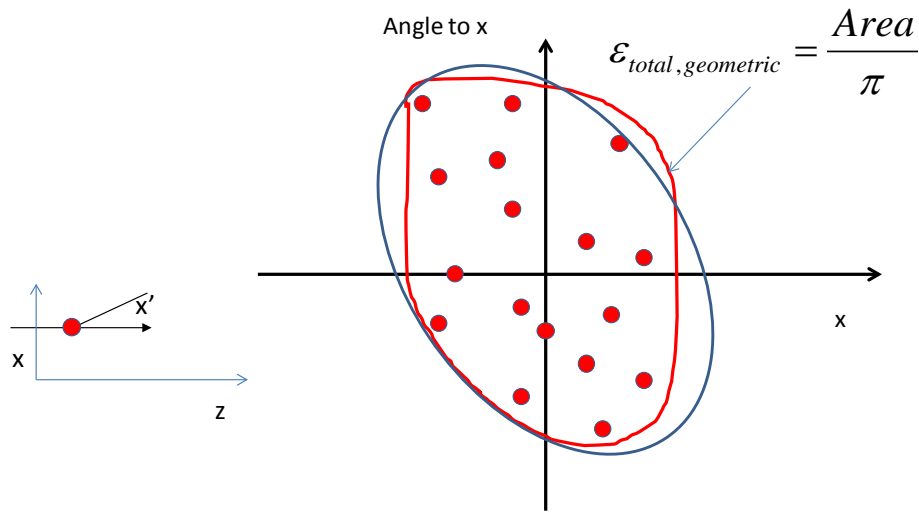

**Fig. 1:** Definition of axes for a beam particle, travelling along the z-axis, at a position *x* and angle *x′*. Many beam particle positions can be plotted in phase space, leading to the definition of a beam emittance.

The solution to Hill's equation describes that a particle orbiting a synchrotron, when measured as a fixed position along the synchrotron, will make the shape of an ellipse in phase space, Thus it is very common to consider the beam from a source also to be enclosed by an ellipse, and therefore the total emittance can be considered to be the area of this ellipse (divided by π).

Very often, the emittance will be quoted as $\varepsilon\,\pi$ mm mrad, i.e. there is the inclusion of a π in the units. Correctly, whether the units contain this *flag* of π or not, the value of $\varepsilon$ is always that corresponding to Eq. (9).

Although the convention of considering an ellipse is retained, ion sources rarely produce a beam that fits well with the definition of an ellipse, and seldom is the beam distribution inside either uniform or Gaussian.

The root mean square (r.m.s.) emittance is a statistical definition of the amount of phase space covered by a beam, and is given by

$$\varepsilon_{\text{r.m.s.}} = \frac{1}{N^2}\left\{\left[N\sum x^2 - \left(\sum x\right)^2\right]\left[N\sum x'^2 - \left(\sum x'\right)^2\right] - \left[N\sum xx' - \sum x \sum x'\right]^2\right\}^{1/2}, \quad (10)$$

$$\varepsilon_{\text{r.m.s.}} = \frac{1}{N}\left(\sum x^2 \sum x'^2 - \left(\sum xx'\right)^2\right)^{1/2}, \quad \overline{x} = \overline{x'} = 0, \quad (11)$$

where the second line is valid when the beam is centred in both position and angle. In this centred beam case, we can also calculate the Twiss parameter of the beam:

$$\gamma_x \varepsilon_{\text{r.m.s.}} = \langle x'^2 \rangle, \qquad \alpha_x \varepsilon_{\text{r.m.s.}} = \langle xx' \rangle, \qquad \beta_x \varepsilon_{\text{r.m.s.}} = \langle x^2 \rangle. \quad (12)$$

The value of the emittance defined above is reduced when the ions are accelerated. In order to express this requirement more clearly, the definition of an emittance that does not change with energy is needed, which is called the normalized emittance. This is normally symbolized by $\varepsilon_n$ and sometimes the symbol $\varepsilon^*$ is especially used to note that the emittance is normalized, 1 r.m.s.

We also consider the concept of the beam intensity normalized by the beam emittance, for which we give the name beam brightness ($B_b$):

$$B_b = \frac{I_b}{\pi^2 \varepsilon_x \varepsilon_y}. \quad (13)$$

Considering the various definitions of the emittance, it is not normally useful to quote a value for a beam brightness, but it is useful for comparing the relative performance of different sources.

### 3.5 Material efficiency

Ion sources take a raw material as an input to produce ions, and this raw material can sometimes be rare, difficult to produce or hazardous. Examples of these can be rare elements (for example, the element iridium) and radioactive isotopes. For ion sources that need these raw inputs, there is therefore a benefit to have a high efficiency of conversion from the input material to the output ion desired.

The simple definition of efficiency is therefore

$$\eta = \frac{N_{\text{ions}}}{N_{\text{atoms}}}. \quad (14)$$

## 3.6 Beam purity

Several applications will require a very pure beam – for example, for hadron therapy – and for many physics experiments – for example, where reaction rates are measured by knowing the incoming beam flux of the ion projectile being studied.

Typically, the ion transportation or acceleration system can play an important role in beam purification, with RF acceleration and spectrometer magnets (and Wien filters) usually allowing ions of the wrong charge-to-mass ratio to be filtered out.

In some cases, this will not be sufficient, or the application is placed close to the source, and therefore the source should already play a role in eliminating unwanted ions. Plasma-based sources are not very well suited to producing clean beams, as they have a large range of electron energies that can ionize any element. Resonant ionization laser ion sources are a good choice, as they are able to use internal excitation of an electron in an atom to an excited state, from which it can be ionized with a low-energy laser (which is not able to ionize other elements that are in their ground states).

Some selectivity is also available from surface ionization sources, where only atoms with low ionization potential will be ionized, but this can still lead to a few different ion types being produced.

## 3.7 Outlet pressure

Most (but not all) sources require a large amount of material to be injected in order to produce ions, and the un-ionized gas or vapour of the material will also exit the source. The gas ejected from the source can lead to breakdowns in both the high-voltage regions of the extraction system and in other high-field regions. The gas can also interact with the beam to cause beam recombination or stripping, which lowers the intensity and can lead to high powers of beam losses for high-intensity sources.

## 3.8 Reliability

As ion sources are the driver of other processes or experiments, it is clear that high reliability is desirable. In general, there will be a trade-off between the other performance factors of the source. For example, high-intensity sources (in particular, if they run at a high duty factor) can suffer from wear due to the ion bombardment of the surfaces, which can also coat other surfaces in the source, thereby limiting the lifetime.

In terms of reliability, any of the final beam parameters of the source being outside defined limits can lead to the source being effectively in fault, depending on the application. Lower than nominal beam intensities, for example, may be tolerated by increasing the length of an experiment, but the ion beam energy not being correct (e.g. due to voltage holding problems) may be equivalent to no beam at all.

Ion source faults can be due to issues with the ion source itself (e.g. failure of a cathode), or to sparking across an insulator, or to an external service to the source (e.g. a high-voltage power supply or failure of the cooling water supply to the source). Faults to the services to the source are often quickly repaired (assuming the spare parts are available), unless the source requires a thermal equilibrium to be reached (e.g. a cathode to be heated to a certain temperature), and the re-establishment of this configuration can take some time. Regarding problems to the source itself, the downtime can be much longer, owing to the requirement to open the vacuum system, and potentially to recondition the source after a change of component (or the complete change of the source).

Like any system, ion source reliability can be characterized by the following metrics:

– Availability, the fraction of the time that a process is working as required.
– Mean time between failures (MTBF), the average time between two faults that need intervention.

- Mean time to repair (MTTR), the average time it takes to repair a fault (weighted with the probability of the fault happening).

In order to increase the availability, there are some general rules that can help to increase the availability of the ion source:

- improve EMI sensitivity to avoid trips due to breakdowns;
- use uninterruptable power supplies to reduce sensitivity to power cuts;
- keep installations clean (inside the vacuum and outside) in order to avoid discharges;
- keep spares, and maintain and test them;
- cold-spare sources (fully built, tested and ready for installation);
- hot-spare sources (two installed sources, with a switch-yard into the beam line, to switch quickly between them).

All of these come at a cost, in both initial investment and operating costs.

### 3.9 Stability

The variation of any of the operating parameters of an ion source can be deemed to lead to an unstable situation for its operation. Although this is most obviously seen if the beam intensity is varying, other beam parameters are, of course, of concern to the system utilizing the beam.

## 4 High-charge-state ion sources

The energy of an ion beam can be increased more effectively by increasing the charge state of the ion. By being more effective, ions can therefore be accelerated in shorter distances, which reduces costs in the construction of the accelerator, and typically will reduce the electrical power required to produce the necessary acceleration.

Ion sources produce higher charge states through a process of stepwise ionization, that is, in a chain from $1+ \to 2+ \to \ldots \to n+$. This requires that two conditions be fulfilled so that the charge state $n+$ can be attained.

1. The ionizing process must have sufficient energy that excitation from $(n-1)+ \to n+$ is possible.
2. The ions are kept in the plasma long enough for them to attain the charge state required.

For the second of these points, and assuming that the ionization process is via electron impacts onto ions, a confinement time ($t_{\text{conf}}$) can be defined, which is the product of the electron density ($n_e$) and the lifetime of the ion in the source ($\tau$), that is,

$$t_{\text{conf}} = n_e \tau. \tag{15}$$

The required charge state can be reached either by having a high-density ion ionizing electrons, or by confining the ions for a sufficiently long time.

## 5 Negative ions

The majority of elements in the periodic table have the ability to attach an additional electron, with the energy released by making the attachment being called the *electron affinity*. The values of the electron affinities for the elements of the first two rows of the periodic table are given in Table 1.

**Table 1:** Electron affinities for the elements of the first two rows of the periodic table.

| Element | Electron affinity (eV) | Element | Electron affinity (eV) |
|---|---|---|---|
| Hydrogen (1) | 0.75 | Carbon (6) | 1.26 |
| Lithium (3) | 0.62 | Oxygen (8) | 1.46 |
| Boron (5) | 0.28 | Fluorine (9) | 3.40 |

Negative ions are of special interest for acceleration systems, owing to the ease with which they can be converted into neutral or positive ions. There are several mechanisms for making this conversion: by stripping using material (in either the solid, gas or plasma form), by photo-detachment or by using a sufficiently strong magnetic field to pull the loosely bound electron off the ion.

The interest in changing the polarity of the ion is because this nonlinear process allows the fast reversal of the force produced by an applied electric and magnetic field (or to cancel it in the case of conversion to a neutral atom). The following are four well-known applications of negative ions:

1. *The tandem accelerator.* Negative ions are accelerated from ground potential to a high-voltage terminal, where they are stripped to positive ions, and accelerated by the opposite electric field to ground potential. The ion energy is two (or more) times the applied voltage.

2. *Ion extraction from cyclotrons.* Negative ions are accelerated in a cyclotron, and stripped into protons, allowing the magnetic field to extract them directly, avoiding complicated extraction geometries.

3. *Charge exchange injection into synchrotrons.* Negative ions are accelerated in a linear accelerator, and overlapped onto circulating positive ions in a synchrotron (due to the opposite bending in a magnetic field). Both beams are passed through a stripping system, emerging as positive ions in a higher-brightness beam.

4. *Magnetic confinement fusion – neutral beam injection.* Negative ions are accelerated, then stripped into a neutral, high-energy (and high-power) beam that can cross the magnetic fields into a magnetic confinement region.

A further application that is far beyond the scope of this chapter is the creation of polarized protons, where the orbital electrons are used to align the angular moment of the proton/nucleus.

Negative ions bring with them further challenges for source engineering and beam transport, including dealing with electron beams co-extracted with the ions, and unwanted gas stripping in the transport lines, but they do bring with them some interesting possibilities for beam instrumentation.

# 6    Summary

In this chapter, the requirements for ion sources and the ways to specify them have been given. In the following chapters, ion sources will be sorted into general types, and will be explained in more detail.